\documentclass[%
preprint,
 amsmath,amssymb,
 aps,
]{revtex4-2}

\usepackage{graphicx}
\usepackage{dcolumn}
\usepackage{bm}

\begin{document}
\title{Dynamically reconfigurable multi-wavelength interferometry}
\author{Leonard Voßgrag}
\email{leonard.vossgrag@imtek.uni-freiburg.de}
\author{Ingo Breunig}
 
\affiliation{Laboratory for Optical Systems, Department of Microsystems Engineering - IMTEK, University of Freiburg, Georges-Koehler-Allee 102, 79110 Freiburg, Germany
}

\author{Annelie Schiller}
\author{Tobias Seyler}
\author{Markus Fratz}
\author{Alexander Bertz}
\author{Daniel Carl}
 
\affiliation{Fraunhofer Institute for Physical Measurement Techniques IPM, Georges-Koehler-Allee 301, 79110 Freiburg, Germany
}

\date{\today}

\begin{abstract}
Single-wavelength interferometry achieves high resolution for smooth surfaces but struggles with rough, industrially relevant ones due to limited unambiguous measuring range and speckle effects. Multi-wavelength interferometry addresses these challenges by using synthetic wavelengths, enabling a balance between extended measurement range and resolution by combining several synthetic wavelengths. This approach holds immense potential for diverse industrial applications, yet it remains largely untapped due to the lack of suitable light sources. Existing solutions are constrained by limited flexibility in synthetic-wavelength generation and slow switching speeds. We demonstrate a light source for multi-wavelength interferometry based on electro-optic single-sideband modulation. It reliably generates synthetic wavelengths with arbitrary values from centimeters to meters and switching times below 30~ms. This breakthrough paves the way for dynamic, reconfigurable multi-wavelength interferometry capable of adapting to complex surfaces and operating efficiently even outside laboratory settings. These capabilities unlock the full potential of multi-wavelength interferometry, offering unprecedented flexibility and speed for industrial and technological applications. 
\end{abstract}
\maketitle

\section{Introduction}

The surface shape of an object can be measured by illuminating it with coherent light of wavelength $\lambda$ and analyzing the interference pattern formed by the superposition of the wave reflected from the surface and a reference wave. For accurate measurements, the surface must be "smooth." Specifically, the interference pattern should be free of speckles, and any abrupt changes in the surface shape must be smaller than $\lambda/2$. Interferometry has emerged as a powerful technique for determining the shape of optically flat surfaces, typically achieving resolutions of $\lambda/100$ or better. In addition to its high resolution, interferometry offers a significant advantage in acquisition speed compared to tactile surface mapping methods.

However, a vast class of functionally rough surfaces cannot be classified as "smooth." These surfaces are of great importance because they are widely encountered in industrial and technological applications, including machined, cast, forged, welded, and 3D-printed components. 
The introduction of multi-wavelength interferometry has enabled the interferometric determination of their shapes\cite{Takeda94}. This approach has been studied for over half a century \cite{Haines65,Hildebrand67,Wyant71,Polhemus73}. It uses coherent light at two wavelengths, $\lambda_1$ and $\lambda_2$. By evaluating and subtracting the resulting phase maps, speckle patterns are suppressed, producing a phase map equivalent to one obtained with coherent light at the synthetic wavelength $\Lambda=c/|\nu_2-\nu_1|$, where $c$ is the speed of light, and the optical frequencies are given by $\nu_{1,2}=c/\lambda_{1,2}$. The synthetic wavelength extends the unambiguous measurement range $\Lambda/2$, by orders of magnitude compared to single-wavelength interferometry, albeit with a proportional decrease in resolution.

This trade-off is addressed by employing multiple synthetic wavelengths through hierarchical unwrapping of phase maps \cite{Wagner00}. The longest synthetic wavelength is chosen to be at least twice the largest deformation of the surface, ensuring an adequate measurement range, while the shortest wavelength determines the resolution. Selecting intermediate wavelengths between these extremes is crucial for accurate phase unwrapping, especially for surfaces with complex deformations or noisy data \cite{Goodman76, Charriere07}. Too few wavelengths risk phase ambiguity, while too many increase measurement time and complexity without significant accuracy gains.

The significant potential of multi-wavelength interferometry lies in its ability to flexibly adjust synthetic wavelengths across the desired range, providing a balance between resolution and unambiguous measurement range. This adaptability, in principle, makes it an invaluable tool for characterizing a wide variety of surfaces in industrial and technological applications. However, despite decades of research \cite{Osten22}, its full potential remains untapped. One contributing factor is the absence of suitable light sources capable of reliably and rapidly altering the synthetic wavelength across several orders of magnitude \cite{Yang18}.

\begin{figure}[!]
    \centering
    \includegraphics[width=170mm]{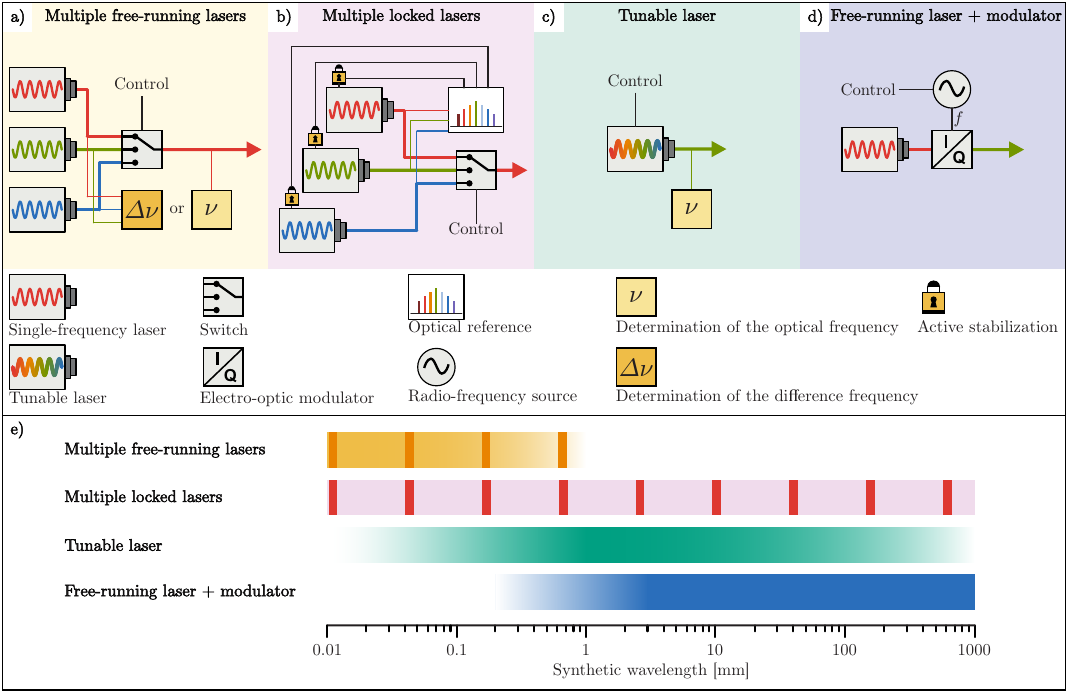}
    \caption{Schematic representations (a-d) of different approaches for generating synthetic wavelengths and their respective ranges (e). The vertical orange and red lines in (e) indicate that this approach provides only a discrete set of synthetic wavelengths.}
    \label{fig:comparison_Lightsource}
\end{figure}

Various methods exist for generating synthetic wavelengths. One commonly used approach relies on multiple free-running lasers (Fig.~\ref{fig:comparison_Lightsource}a), each operating at its specific optical frequency \cite{Fratz19,Weimann15,MeinersHagen04}. A switching mechanism ensures that interferograms are recorded sequentially at each optical frequency. The synthetic wavelengths are then derived from the corresponding difference frequencies. In this method, the lasers are selected to produce synthetic wavelengths tailored to a particular surface to be measured, which inherently limits flexibility. Additionally, accurate determination requires measuring the optical frequencies of the free-running lasers, their individual frequencies, or their differences. The uncertainty in these optical frequencies restricts the practical synthetic wavelengths of such devices to millimeter-scale values or smaller. Despite of its limitations, this approach has successfully found a way out of the lab into industrial applications \cite{Fratz21}. 

The range of synthetic wavelengths can be significantly extended by locking the emission frequencies of individual lasers to a suitable reference, such as a frequency comb (Fig.\ref{fig:comparison_Lightsource}b) \cite{Wang15,MeinersHagen09}. In this configuration, the smallest difference frequency is dictated by the free spectral range of the reference. For example, a frequency comb with a 100~MHz free spectral range allows for synthetic wavelengths reaching several meters. However, despite the considerably increased complexity of this approach, it remains fundamentally limited. Since it relies on individual lasers, it still offers only a discrete set of synthetic wavelengths, which must be carefully optimized for specific surface measurements.

To address this limitation, a tunable laser can be used instead of multiple lasers (Fig.~\ref{fig:comparison_Lightsource}c) \cite{Fischer95,Seyler22,Groger23}. In this approach, the synthetic wavelength can be adjusted within a range determined by two factors: the tuning range of the laser, which sets the minimum value, and the tuning step size, which determines the maximum value. For mode-hop-free tunable lasers, the step size is limited by the uncertainty in the optical frequency, enabling synthetic wavelengths to extend to meter scales. While this method offers a high degree of flexibility, reliably tuning the laser across a broad range requires precise frequency tracking and a control loop. Despite its versatility, the approach is constrained by relatively slow tuning speeds.

In this paper, we demonstrate an approach that allows for reliably and rapidly (at timescales below 30~ms) varying the synthetic wavelength arbitrarily across multiple orders of magnitude which are prerequisites for dynamically reconfigurable multi-wavelength interferometry. The approach is based on the following idea: single-frequency light at the optical frequency $\nu$ is converted by a frequency shifter, which is driven at the radio frequency $f$, to produce light at the new frequency $\nu+f$ (Fig.~\ref{fig:comparison_Lightsource}c). Here, the synthetic wavelength is given by $c/f$. Thus, no additional monitoring of the optical frequency is necessary. The application of radio-frequency controlled frequency shifters for multi-wavelength interferometry was demonstrated with acousto-optic modulators \cite{Vossgrag24}. However, these devices do not operate in a wide radio-frequency range. Thus, the flexibility was very limited. In this contribution, we use an electro-optic single-sideband modulator as frequency shifter. These devices can be operated from 0.1~GHz up to the 100~GHz-range \cite{Xu20,Wang18,Xu22}. Thus, synthetic wavelengths can be adjusted to any value between millimeters and meters.

In our proof-of-concept study, we demonstrate a generator for synthetic wavelengths based on single-sideband modulation. This device is characterized regarding spectral purity, flexibility of the synthetic wavelengths and tuning speed. We show its application for multi-wavelength-holography. Two different samples - one machine-milled part made of metal, one made out of plastic via injection-molding - are investigated with different cascades of synthetic wavelengths.

\section{Prerequisites for modulator-driven generation of synthetic wavelengths\label{sec:Prerequisites}}
The generation of synthetic wavelengths with a radio-frequency driven modulator is based on the following idealizations. The output spectrum contains only the shifted optical frequency component at $\nu+f$, where $\nu$ is the original optical frequency and $f$ is the radio frequency. No additional spectral components, such as sidebands or noise, are present. The synthetic wavelength is precisely given by $\Lambda=c/f$, while $f$ is variable over a wide frequency range. In the following, we evaluate how closely electro-optic single-sideband modulation aligns with these idealizations.

We start by examining the spectral purity of the output spectrum. Single-sideband modulation relies on the interferometric combination of four phase modulators, each producing multiple sidebands. By precisely tuning the phase relationships between the outputs of these modulators, the power of the sideband at the desired frequency $\nu+f$ is enhanced to exceed that of all other spectral components by several orders of magnitude \cite{Chen21}. Commercially available devices achieve side-mode suppression ratios exceeding 20~dB.

\begin{figure}[t]
    \centering
    \includegraphics[width=170mm]{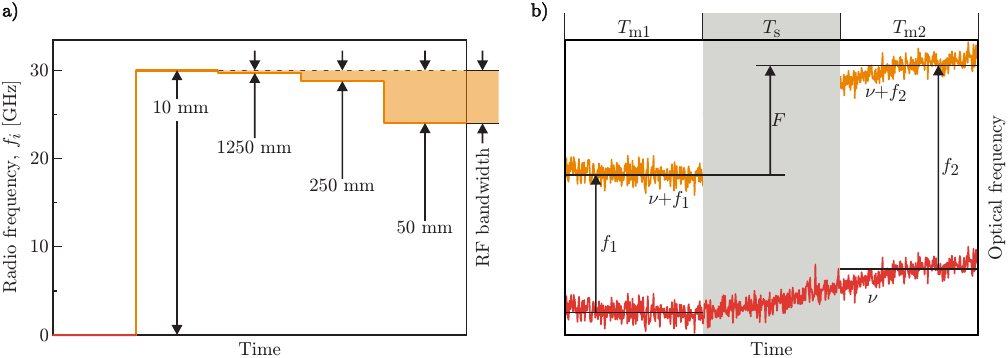}
    \caption{a) Adjusting the synthetic wavelength between 10 and 1250~mm in four steps requires radio frequencies in the range of 24 to 30~GHz. b) Temporally varying frequencies $\nu$ of laser light and sidebands $\nu+f_{1,2}$ generated by a radio-frequency-controlled frequency shifter. Interferograms at $\nu+f_{1,2}$ are recorded during $T_{\rm{m1,2}}$, while $T_{\rm{s}}$ represents the switching time for changing the radio frequency from $f_1$ to $f_2$. The synthetic frequency $F$ may differ from the exact value $f_2-f_1$.}
    \label{fig:drift}
\end{figure}

Strictly following the idealized relation $\Lambda=c/f$ indicates that the required radio-frequency bandwidth is given by $B=f_{\rm max}-f_{\rm min}=c/\Lambda_{\rm min}-c/\Lambda_{\rm max}$. Since our goal is to vary the synthetic wavelength over orders of magnitude, we can assume $\Lambda_{\rm max}\gg\Lambda_{\rm min}$ and consequently $B\approx c/\Lambda_{\rm min}$. If we want to determine phase maps between 10 and 1250~mm synthetic wavelengths in four steps, we need the radio frequencies 0, 0.24, 1.2, 6 and 30~GHz, i.e. a bandwidth close to 30~GHz. However, the same synthetic wavelengths can be generated by using 0, 24, 28.8, 19.87 and 30~GHz as visualized in Fig.~\ref{fig:drift}a. Thus, if we determine the synthetic wavelength from $\Lambda=c/|f_2-f_1|$, the required radio-frequency bandwidth can be significantly decreased.

However, this is still an idealization that holds true only if the laser frequency $\nu$ remains constant. In practice, however, lasers exhibit short-term fluctuations, defined by their linewidth, and long-term drifts, as shown in Fig.~\ref{fig:drift}b. To determine a phase map at a specific synthetic wavelength, the radio frequency is set to $f_1$ for a duration $T_{\rm m1}$ to capture the first set of interferograms. The frequency is then switched to $f_2$ acquiring a time $T_{\rm s}$, followed by capturing the second set of interferograms over $T_{\rm m2}$. The total time $T_{\rm m1}+T_{\rm s}+T_{\rm m2}$ must be much shorter than the typical timescale of long-term drift to minimize synthetic wavelength uncertainty, which is then determined only by the laser’s linewidth. Using a single-frequency laser with a MHz-level linewidth, one can reliably determine synthetic wavelengths up to meters, provided the total duration for capturing and switching remains on the order of seconds.

\section{Synthetic-wavelength generator}
\subsection{Experimental implementation}
\begin{figure}[t]
\centering\includegraphics[width=170mm]{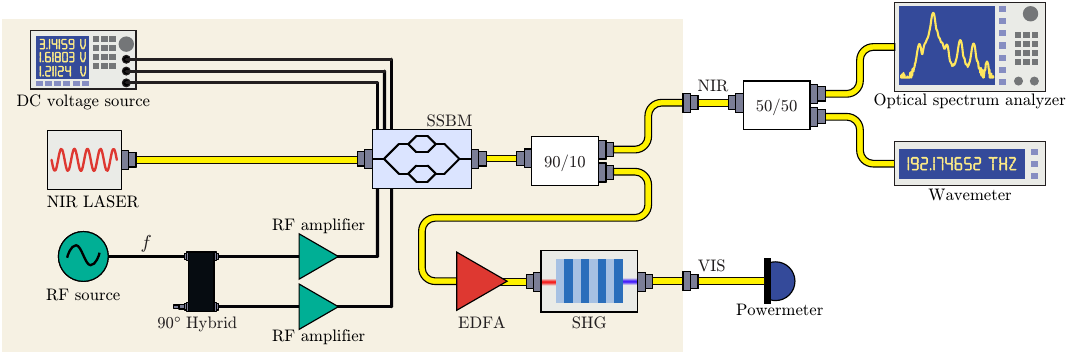}
\caption{Schematic setup of the synthetic wavelength-generator comprising a near-infrared (NIR) laser single-sideband modulator (SSBM) driven at the radio frequency $f$, an erbium-doped amplifier (EDFA) and a frequency doubler (SHG). The NIR output serves for spectral characterization whereas the VIS output is connected with a powermeter.}
\label{fig_setup_char}
\end{figure}

Figure \ref{fig_setup_char} shows the setup of the synthetic-wavelength generator. Near infrared light at 1560~nm wavelength with less than 1~MHz linewidth is provided by a  fiber laser (Koheras BasiK E15) and shifted in frequency by a single sideband modulator (Exail MXIQER LN30). The modulator is driven by two radio-frequency signals at the frequency $f$ phase-shifted by $\pi/2$. They are delivered by the combination of a tunable signal generator (Anritsu MG3696A), a hybrid coupler (Sigatek SQ20509) and two amplifiers (Exail DR-VE-10-MO). Furthermore, three DC voltages are fed into the modulator in order to optimize the side-mode suppression. 

The interferometer (Fraunhofer IPM,HoloTop NX NIR) available for this study requires visible light around 780~nm wavelength. Thus, 90~\% of the the near-infrared light converted into the visible spectral range using an optical amplifier (IPG, EAD-3-C-PM) and a frequency doubler (HC Photonics, PMC2307050016). The rest of the near-infrared light is used for characterization.

\subsection{Characterization}
In order to be useful for multi-wavelength interferometry, the output power of the synthetic-wavelength generator should be in the milliwatt range at 780~nm wavelength. As discussed above, it needs to provide spectrally pure light. The values for $\Lambda$ should be adjustable to arbitrary values over two orders of magnitude with switching times far below one second.

The output power at 780~nm wavelength is measured to be 2 mW when the pump laser emits 15~mW at 1560~nm wavelength and the erbium-doped amplifier is set to 200~mW output power. 

\begin{figure}[t]
\centering\includegraphics[width=170mm]{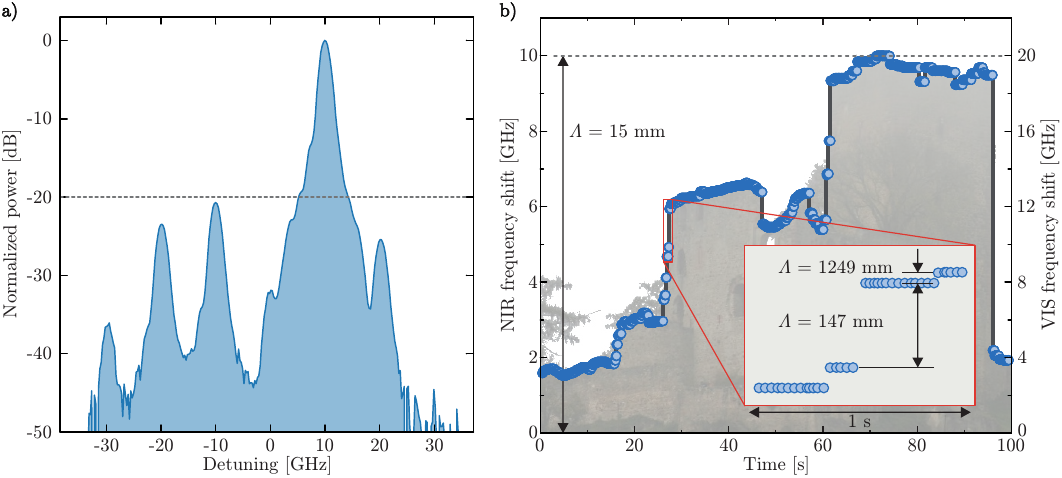}
\caption{a) Output spectrum in the near infrared when the modulator is driven at 10~GHz radio frequency. b) Measured frequency shift in the near infrared and corresponding values in the visible as function of time. The shape of the curve resembles the contour of the remains of the "Hochburg Emmendingen" near Freiburg, Germany. Some characteristic frequency shifts are denoted with respective values for the respective synthetic wavelengths $\Lambda$.}
\label{fig:Spectrum_Tuning}
\end{figure}

The spectral purity is determined with an optical spectrum analyzer (OSA,Yokogawa AD6730). Fig.~\ref{fig:Spectrum_Tuning}a exemplary shows the near-infrared output spectrum when the modulator is driven at 10~GHz radio frequency. The side-mode suppression is better than 20 dB. Since the conversion efficiency of frequency doubling scales quadratically with input power, we assume that the side-mode suppression in the visible spectral range is better than 40 dB which is in line with recently published data \cite{Sabatti24}. 

In order to demonstrate the flexibility of the synthetic-wavelength generator, we vary the radio frequency $f$ over a period of 100~s in such a way that the function $f(t)$ resembles the contour of the remains of the "Hochburg Emmendingen" near Freiburg, Germany. Simultaneously, we record the near infrared output wavelength with a wavemeter (HighFinesse, WS7-60) and determine the corresponding frequency shift from the initial laser frequency. Fig.~\ref{fig:Spectrum_Tuning}b shows the near infrared frequency shift as well as the doubled value that is expected for the visible output. We do not observe any significant difference between the target values given by $f(t)$ and the measured frequency shifts. We have reliably generated frequency shifts between the the hundred MHz region up to 20~GHz which is given by the maximum frequency of the hybrid coupler. This result shows that we can vary the synthetic wavelength to arbitrary values between 15 mm and more than one meter. The 1~s long zoom in Fig.~\ref{fig:Spectrum_Tuning} shows that the switching takes less than 30~ms which is limited by the measurement period of the wavemeter. 

The performance of the synthetic-wavelength generator regarding output power, spectral purity, flexibility and switching time meets all abovementioned requirements for multi-wavelength interferometry. Furthermore, we do not observe any significant changes in its performance over several hours without any active stabilization.

However, these values do not represent the limit of this approach. Erbium-doped amplifiers are capable of delivering output powers in the Watt range, which is ten times higher than the power levels used in our experiment. At such power levels, the frequency-doubling stage could produce visible light with an output of approximately 100 mW. Additionally, we are witnessing significant advancements in the performance of electro-optic devices, driven by the adoption of thin-film lithium niobate \cite{DiZhu21}. This technology enables single-sideband modulators to operate at radio frequencies exceeding 100 GHz. Furthermore, by changing the DC voltages at the modulator, the output of the single-sideband modulator can be switched from $\nu+f$ to $\nu-f$. Combining both strategies, synthetic wavelengths below 1~mm are achievable. Regarding switching speed, the limit is set by the radio-frequency signal generator. Here, values below 15~\textmu s are typical \cite{Banerjee20}, i.e. several orders of magnitude below the limit of the wavemeter used in our experiment.

\section{Determination of surface shapes}
\begin{figure}[t]
\centering
\includegraphics[width=170mm]{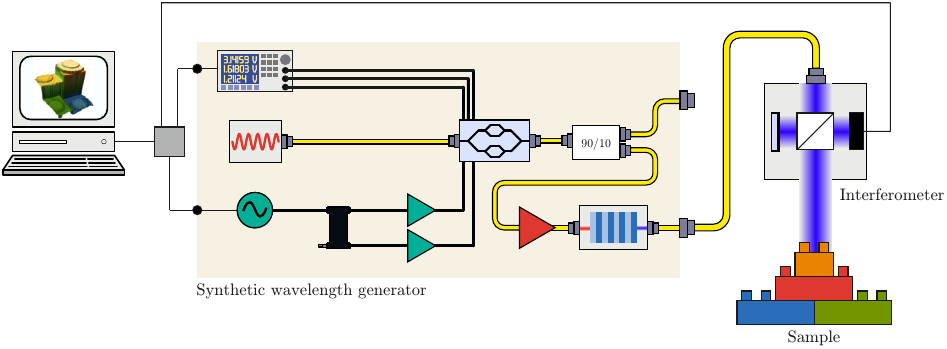}
\caption{Setup for holographic measurement. The synthetic-wavelength-generator is connected to the interferometer via the 780-nm-wavelength port. A connected computer receives the images from the camera for computing the phase-maps and sets the bias voltages of the single-sideband modulator and the desired microwave-frequencies $f$ for the corresponding synthetic wavelength $\Lambda$.}
\label{fig_setup}
\end{figure}

In order to showcase the applicability of the synthetic-wavelength generator demonstrated above, we connect the visible light output to a commercially available holographic measurement sensor (Fraunhofer IPM, Holotop NX NIR), as illustrated in Fig.\ref{fig_setup}. This sensor incorporates a Mach-Zehnder-type interferometer. To determine the phase maps, a three-step phase-shifting method is employed \cite{Greivenkamp84,Cai04}. Consequently, three interferograms are recorded for each radio frequency $f$, requiring approximately 100~ms. This duration is determined by the motion of a piezo actuator used for phase shifting and the camera's exposure time. For a series of four synthetic wavelengths, as shown in Fig.\ref{fig:drift}a, all necessary interferograms are captured in less than a second. The accompanying software (Fraunhofer IPM, Holo-Software) handles phase reconstruction, hierarchical unwrapping, and numerical propagation. Additional details about the sensor head can be found in Ref.~\cite{Stevanovic21}

\begin{figure}[t]
\centering
\includegraphics[width=170mm]{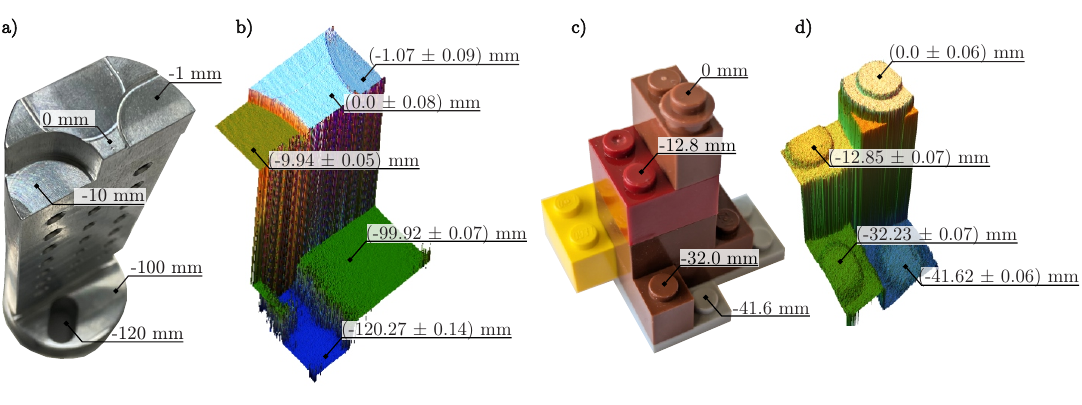}
\caption{Photographs (a and c) of the objects investigated and corresponding interferometrically determined surface shapes (b and d). The photographs comprise height values given by the manufacturer. The corresponding values in b and d are means and respective standard deviations over  $200\times200$ pixel subsections.}
\label{fig_holograms}
\end{figure}

To show the flexibility of the synthetic-wavelength generator, we use two different samples of different size and material. Our first sample is a machine-milled piece of metal with several sub-surfaces at distances between 1 and 120~mm (see Fig.~\ref{fig_holograms}a). Thus, the maximum synthetic wavelength has to be larger than 240~mm. Since the smallest synthetic wavelength is 15~mm, we choose 15, 75 and 375~mm as synthetic wavelengths, i.e. two factor-of-five steps. Here, the total reconstruction time was 1.3~s including 0.7~s for capturing 12 interferograms with 9 million pixels each and 0.6~s for their numerical evaluation. Figure~\ref{fig_holograms}b shows the reconstructed surface shape with the respective heights of the characteristic sub-surfaces. It is apparent that the standard deviation of these values are even smaller than a hundredth of the smallest synthetic wavelength. Furthermore, all mean values nicely agree with the nominal ones.

The second sample is composed of multiple commercially available toy-building-blocks, made of plastic via injection molding (see Fig.~\ref{fig_holograms}c). The total height of the object is approximately 45~mm. Furthermore, we want to maintain the smallest wavelength of 15~mm for the highest resolution. In order to minimize the number of interferograms, we chose 15 and 150~mm, i.e. only one factor-of-ten step. Here, the total reconstruction time was reduces to 1.0~s. Figure~\ref{fig_holograms}d shows the reconstructed surface shape. Also here, the determined values nicely agree with the expected ones.

Our experiments show that the synthetic-wavelength generator described above can be reliably applied for multi-wavelength interferometry. The total reconstruction time is on the level of a seconds for three synthetic wavelengths. This duration can be further decreased. Applying spatial phase shifting \cite{Ichioka72} rather than temporal phase shifting reduces the capture time to the time that it takes to expose the camera chip and to save the data. Commercially available devices provide this on the level below 10~ms. Thus, with improved hardware, data capture could be faster by a factor of ten.

\section{Conclusion}
In conclusion, we have proposed and demonstrated a synthetic-wavelength generator for multi-wavelength interferometry based on electro-optical single-sideband modulation. It provides a variation of the synthetic wavelength to arbitrary values between 15 and 1500~mm with switching times below 30~ms. This combination of flexibility and speed paves the way for dynamically reconfigurable interferometric measurements. The actual values for the synthetic wavelength can be specifically optimized to any reflecting object without changing the hardware of the light source. We have demonstrated that surfaces with deformations larger than 100~mm can be reliably reconstructed with uncertainties of the level of 0.1~mm. 

We have discussed that the performance of the light source can be significantly improved regarding output power, smallest synthetic wavelength and data-capture time. Such synthetic-wavelength generators pave the way for reconstructing large-area surfaces deformations in the meter range and resolution below 10~\textmu m. Furthermore, with data-capture times below 10~ms, even temporally varying surfaces can be investigated. 

The concept is based on integrating a laser with a single-sideband modulator, an erbium-doped amplifier, and a frequency doubler. The latter three components have already been implemented using thin-film lithium niobate \cite{Boes23,Xue24,Zhang21}. As a result, they can be integrated into a single chip, significantly reducing the size of the optical setup and enhancing its robustness.

We believe that the capabilities offered by this concept will greatly expand the range of applications for multi-wavelength interferometry. It will not only benefit fundamental scientific research but, thanks to its robust components, also enable applications beyond the laboratory environment.

\section*{Funding}
German Federal Ministry of Education and Research, Research Program Quantum Systems, 13N16774

\section*{Disclosures}
\noindent Authors A.S., T.S., M.F., A.B. and D.C. are employed by the Fraunhofer Institute for Physical Measurement Techniques, IPM, which commercially supplies the interferometric sensor utilized in this work.

\section*{Data availability} Data underlying the results presented in this paper can be obtained from the authors upon reasonable request.

\bibliography{Bibo.bib}

\end{document}